\documentclass[letterpaper]{article} % DO NOT CHANGE THIS
\usepackage{aaai2026}  % DO NOT CHANGE THIS
\usepackage{times}  % DO NOT CHANGE THIS
\usepackage{helvet}  % DO NOT CHANGE THIS
\usepackage{courier}  % DO NOT CHANGE THIS
\usepackage[hyphens]{url}  % DO NOT CHANGE THIS
\usepackage{graphicx} % DO NOT CHANGE THIS
\usepackage{natbib}  % DO NOT CHANGE THIS AND DO NOT ADD ANY OPTIONS TO IT
\usepackage{caption} % DO NOT CHANGE THIS AND DO NOT ADD ANY OPTIONS TO IT
\usepackage{amsfonts}
\usepackage{amsmath}
\usepackage{multirow}

\usepackage{algorithm}
\usepackage{algorithmic}

\usepackage{newfloat}
\usepackage{listings}
\DeclareCaptionStyle{ruled}{labelfont=normalfont,labelsep=colon,strut=off} % DO NOT CHANGE THIS
\floatstyle{ruled}
\newfloat{listing}{tb}{lst}{}
\floatname{listing}{Listing}
\title{Graph Neural Field with Spatial-Correlation Augmentation for HRTF Personalization}
\author {
    De Hu\textsuperscript{\rm 1}\thanks{Corresponding Author}, Junsheng Hu\textsuperscript{\rm 1}, CuiCui Jiang\textsuperscript{\rm 1}
}
\affiliations {
    \textsuperscript{\rm 1}College of Computer Science, Inner Mongolia University, China\\
    cshood@imu.edu.cn, hujunsheng@mail.imu.edu.cn, 
    jiangcuicui@mail.imu.edu.cn
}

\begin{document}

\maketitle

\begin{abstract}
To achieve immersive spatial audio rendering on VR/AR devices, high-quality Head-Related Transfer Functions (HRTFs) are essential. 
In general, HRTFs are subject-dependent and position-dependent, and their measurement is time-consuming and tedious.
To address this challenge, we propose the Graph Neural Field with Spatial-Correlation Augmentation (GraphNF-SCA) for HRTF personalization, which can be used to generate individual HRTFs for unseen subjects.
The  GraphNF-SCA consists of three key components: an HRTF personalization (HRTF-P) module, an HRTF upsampling (HRTF-U) module, and a fine-tuning stage.
In the HRTF-P module, we predict HRTFs of the target subject via the Graph Neural Network (GNN) with an encoder-decoder architecture, where the encoder extracts universal features and the decoder incorporates the target-relevant features and produces individualized HRTFs.
The HRTF-U module employs another GNN to model spatial correlations across HRTFs. This module is fine-tuned using the output of the HRTF-P module, thereby enhancing the spatial consistency of the predicted HRTFs.
Unlike existing methods that estimate individual HRTFs position-by-position without spatial correlation modeling, the GraphNF-SCA effectively leverages inherent spatial correlations across HRTFs to enhance the performance of HRTF personalization.
Experimental results demonstrate that the GraphNF-SCA achieves state-of-the-art results \footnote{\url{https://github.com/hu-junsheng/GraphNF-SCA.}}.
\end{abstract}

\vspace{-0.5cm}
\section{Introduction}
Spatial audio rendering finds various applications, including virtual reality (VR) \cite{johansson2019vr, zotter2019ambisonics}, augmented reality (AR) \cite{sundareswaran20033d, yang2019audio}, teleconferencing \cite{8013136}, and hearing assistive devices \cite{du2023wearable, vickers2021involving}. 
To generate high-quality immersive audio over headphones, head-related impulse responses (HRIRs), known as head-related transfer functions (HRTFs) in the frequency domain, are often used to simulate the spatial filtering effects, as sound travels from its source to each ear. 
When virtual source positions are given, binaural signals can be synthesized by the direct convolution of HRIRs and source signals.

\begin{figure}[htbp]
\centerline{\includegraphics[width=0.9\columnwidth]{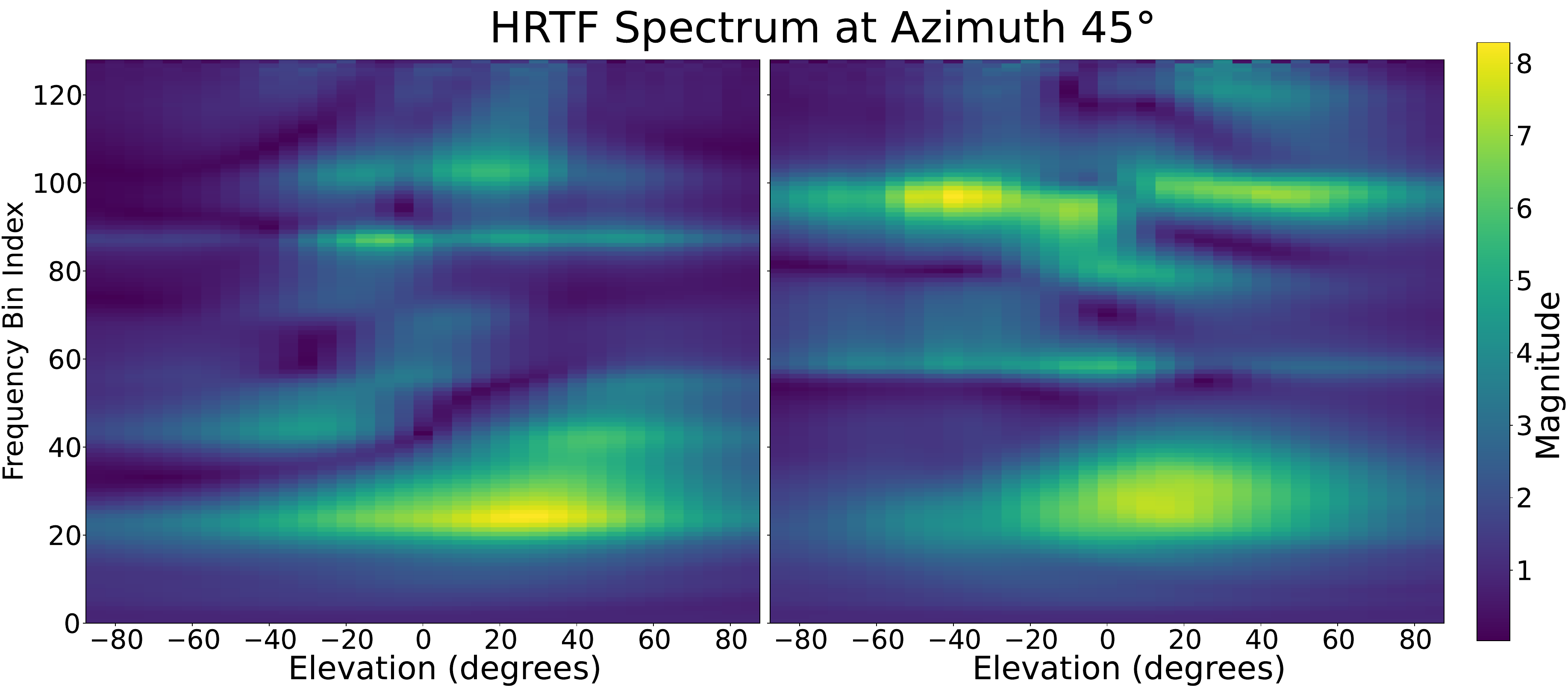}}
\caption{Illustration of position-dependent and subject-dependent nature of HRTFs: The magnitude of HRTFs corresponding to subjects PP1 (left) and PP3 (right) at an azimuth angle of 45°, obtained from the HUTUBS database \cite{brinkmann2019cross}.}
\label{fig:1}
\end{figure}

In general, accurate HRTFs can be obtained by practical measurement. For example, the CIPIC database \cite{algazi2001cipic} collected HRTF data from subjects in multiple spatial orientations in an anechoic chamber, by placing miniature microphones in the subjects' ear canals and utilizing a robotic arm to control the loudspeaker. However, acquiring HRTFs through physical measurements is time-consuming and labor-intensive, thereby limiting the scale of existing HRTF datasets \cite{sridhar2017database, bomhardt2016high, carpentier2014measurement}.
To alleviate this issue, {\ttfamily{HRTF upsampling}} is explored to generate dense HRTFs from sparsely measured HRTFs \cite{porschmann2019directional}.
Furthermore, as depicted in Figure \ref{fig:1}, HRTFs are different for each subject due to their close relationship with anatomical traits (e.g., torso, head, and pinnae). 
As a result, the use of existing HRTF datasets for new users may lead to anatomical mismatches, which in turn degrade the spatial audio experience \cite{simon2016perceptual, jenny2020usability}.
To avoid re-measurement of HRTFs for each new user, {\ttfamily{HRTF personalization}} is employed to generate HRTFs matching the individual's anatomy via existing datasets. 

\vspace{-0.2cm}

\vspace{-0.7cm}
\subsection{Related Works}
\subsubsection{HRTF Personalization (HRTF-P)}
To generate individual HRTFs for unseen subjects, several HRTF-P methods have been proposed.
The most straightforward approaches \cite{zhou2021predictability, teng2023individualized, ko2023prtfnet} were to directly learn the mapping from anatomical features to individual HRTFs.
For example, 3D head scans were utilized as anthropometric features in \cite{zhou2021predictability}, while anthropometric parameters (about head, torso, and pinna ) were adopted in \cite{teng2023individualized, ko2023prtfnet}.
Nevertheless, learning direct anthropometric-to-HRTF mappings faces fundamental challenges in data-scarce scenarios, due to the fact that the underlying wave-anatomy interactions (e.g., diffraction, scattering, reflection, and resonance)  exhibit strong nonlinearity \cite{takemoto2012mechanism, algazi2002approximating, 10836943}. 
An alternative scheme is to estimate the individual HRTF without direct mapping, but by 
retrieving and synthesizing anatomically similar subjects' HRTFs from a database \cite{10889481, 10448477, 1285855, geronazzo2019applying}.
The simplest retrieval-based approaches \cite{1285855, geronazzo2019applying, katz2012perceptually} output a best-match HRTF by searching the subject with the most similar anthropometric features in a dataset. In such a manner, it is unable to generalize the relationship between the input and the HRTF, limiting the performance in small-scale datasets \cite{brinkmann2019cross, pelzer2020head}.
Instead of using individual anatomy, Masuyama et al.~\cite{10889481} proposed a retrieval-augmented neural field (RANF) to carry out HRTF personalization, in which the interaural time difference (ITD) features were employed to retrieve multiple subjects from a given HRTF dataset.
This method requires measuring the target user's HRTFs from only 3-5 directions to calculate ITD features, dispensing with time-consuming anthropometric feature acquisition.
Unfortunately, like other HRTF-P approaches mentioned above, it cannot generate personalized HRTFs at non-sampled positions (that are outside the given dataset).
Furthermore, the estimation accuracy of personalized HRTFs is still limited due to the complicated wave-anatomy interactions.

\vspace{-0.28cm}
\subsubsection{HRTF Upsampling (HRTF-U)}

To generate HRTFs at unsampled positions, HRTF-U can be achieved based on traditional spatial interpolation methods, such as barycentric interpolation \cite{poirier2018anaglyph, hartung1999comparison} and spherical harmonic interpolation \cite{evans1998analyzing, arend2021assessing, engel2022assessing}. 
These methods achieve high interpolation accuracy if the measurement points in the adopted HRTF dataset are distributed densely and evenly. This is because HRTFs tend to be similar for points that are very close to each other, as shown in Figure \ref{fig:1}.
However, existing databases are typically limited in scale, leading to significant performance degradation in conventional interpolation methods. 
With the rapid development of deep learning, data-driven HRTF interpolations have gained increasing attention \cite{14hogg2024hrtf,15gebru2021implicit}. 
Recently, Lee et al. \cite{20lee2023global} interpolated HRTFs by mapping those HRTFs from neighboring points to target points via the neural network that was built upon Feature-wise Linear Modulation (FiLM). Since the input size is required to be fixed in such an architecture, a repeated sampling strategy was used to select a fixed number of neighbors around the target point. 
Alternatively, the state-of-the-art work \cite{hu} introduced a dual-graph attention network with variable input sizes, which shows the strong capability of graph neural networks in modeling spatial correlation among HRTFs.
It can theoretically predict subject-specific HRTFs at any position, but cannot personalize them for unseen subjects.

\vspace{-0.2cm}
\subsubsection{Comparison and Analysis}
As discussed above, the existing HRTF generation approaches can be divided into two classes, i.e., HRTF-P and HRTF-U. 
For a given HRTF dataset containing position-specific and subject-specific measurements, HRTF-P estimates personalized HRTFs for arbitrary subjects at sampled positions, while HRTF-U predicts HRTFs at arbitrary positions for known subjects.
By comparison, HRTF-U often performs well since it focuses more on the spatial correlation among HRTFs. In contrast, HRTF-P demonstrates constrained performance due to the inherent complexity of modeling wave-anatomy interactions.
Until now, to the best of our knowledge,  HRTF-P has always been modeled independently from HRTF-U, which may fail to capture the geometric relationship among personalized HRTFs.
\textit{This naturally raises a critical question: Could developing a collaborative mechanism between HRTF-P and HRTF-U enhance the HRTF-P performance?}

\vspace{-0.2cm}
\subsection{Contribution}

Motivated by the above problem, we propose a novel HRTF-P framework (termed  GraphNF-SCA) in this work. The contributions are threefold. First, we develop a deep learning model for HRTF-P via graph neural networks (GNNs), which employs an encoder to extract universal HRTF features and then integrates target subject-specific characteristics through a decoder to generate personalized HRTFs.
Second, we construct an HRTF-U module based on another GNN, which comprehensively learns the geometric correlations of HRTFs across spatial positions.
Third, and most critically,  we feed the output of the HRTF-P module into the fine-tuned HRTF-U module, which reinforces the initially position-independent personalized HRTFs by leveraging spatial correlations.
Experimental results demonstrate that the estimation accuracy of spatial-correlation augmented HRTFs is significantly higher than that of position-independent baseline methods.
These findings validate our initial hypothesis that integrating HRTF-U can effectively enhance HRTF-P performance.

\vspace{-0.2cm}
\section{Preliminaries}
\subsection{Problem Statement}
Our objective is to generate personalized HRTFs for unseen subjects, using a small-scale dataset with subject-specific and position-specific HRTF measurements.
Due to the fact that the phase of HRTFs can be effectively reconstructed using the minimum-phase approximation \cite{cuevas20193d, 14hogg2024hrtf, 10448477}, this work focuses only on the magnitude of HRTFs.
Let $\mathbf{H}_s^d \in \mathbb{R}^{2K}$ be the HRTF magnitude of subject $s$ at direction $d = (\theta, \varphi)$, where $\theta \in [0, 2\pi)$ and $\varphi \in [-\pi/2, \pi/2]$ are the azimuth angle and the elevation angle, respectively, and $K$ is the number of frequency bins per ear. 
For a given HRTF dataset, we have $s \in \mathcal{S}$ and $d \in \mathcal{D}$ with $\mathcal{S}$ and $\mathcal{D}$ being the subject set and the direction set, respectively.

For an unseen subject $\hat{s} \notin \mathcal{S}$, HRTF-P can be carried out in a retrieval-augmented manner, by learning a mapping
\begin{equation} \label{mapp_hrtf_p}
\Phi_p\left( \mathbf{H}_{s\in \mathcal{N}_{\hat{s}}}^{d} \right) \mapsto  \mathbf{H}_{\hat s}^{d},\; \hat s \notin \mathcal{S}
\end{equation}
where $\mathcal{N}_{\hat{s}}$ is the neighboring set of subject ${\hat{s}}$, which can be acquired by retrieving subjects that have an anatomical feature similar to the target subject, i.e.,
\begin{equation}\label{N_s}
\mathcal{N}_{\hat{s}} = \left\{ s \mid \left \| a_s - a_{\hat s} \right \|_2  < \delta_s \right\}
\end{equation}
where $a_s$ is the subject-relevant characteristic of subject $s$, and $||\cdot||_2$ represents the $l_2$ norm. In the state-of-the-art work \cite{10889481}, $a_s$ is determined by the ITD (if the ITDs of the target subject are measured at a small number of directions in advance) rather than human anatomy. 

For an unmeasured spatial position $\hat d \notin \mathcal{D}$,  recent advances \cite{hu, 20lee2023global} in HRTF-U are also moving toward using the retrieval-augmented strategy to learn the following mapping
\begin{equation} \label{mapp_hrtf_u}
\Phi_u\left( \mathbf{H}_s^{d \in \mathcal{N}_{\hat{d}}} \right)  \mapsto \mathbf{H}_s^{\hat{d}},\; \hat d \notin \mathcal{D} 
\end{equation}
where $\mathcal{N}_{\hat{d}}$ is the neighboring set of point ${\hat{d}}$, which can be obtained by retrieving the points near the target point, i.e.,
\begin{equation}\label{N_d}
\mathcal{N}_{\hat{d}} = \left\{ d \mid \| d - \hat d \|_2 < \delta_d \right\}
\end{equation}
where $\delta_d > 0$ is a proper threshold.

Note that HRTF-P and HRTF-U in (\ref{mapp_hrtf_p}) and (\ref{mapp_hrtf_u}) are modeled independently, where the former exhibits the effect of human anatomy on HRTFs while the latter reflects the effect of source directions on HRTFs. 
\textit{Typically, mapping (\ref{mapp_hrtf_u})  can be learned more easily, whereas mapping (\ref{mapp_hrtf_p}) presents challenges due to complicated wave-anatomy interactions.} To this end, we construct a novel mapping $\Psi_p$:
\vspace{-0.1cm}
\begin{equation}
\Psi_p: 
\boxed{
\tilde\Phi_u\left(\Phi_p( \mathbf{H}_{s\in \mathcal{N}_{\hat{s}}}^{d} ) \!\mapsto \! \mathbf{H}_{\hat s}^{d}, d\in \mathcal{N}_{{d}}\right) \!\mapsto \!\mathbf{H}_{\hat{s}}^{ d}, \;\hat s \notin \mathcal{S}
}
\end{equation}
where $\tilde\Phi_u$ is the fine-tuning version of $\Phi_u$, which is used to reinforce $\Phi_p$ by taking advantage of the spatial relationship among $\mathbf{H}_{\hat s}^{d \in \mathcal{N}_{{d}}}$.

\vspace{-0.3cm}
\section{Method}
In this work, the graph neural network (GNN) is adopted, due to the fact that
\begin{itemize}
    \item GNNs are well suited for modeling the relationship between nodes based on their local similarities, e.g., anthropometric features in (\ref{N_s}) and spatial distances in (\ref{N_d}). 
    %In HRTF upsampling tasks, the smaller the distance between two positions, the more similarity between their HRTFs (e.g., \ref{fig:1}). Analogously, the more similar the anthropometric feature between two subjects, the more similarity between their HRTFs.

    \item GNN supports inputs with non-fixed sizes, which allows for retrieving sets with flexible sizes in (\ref{N_s}) and (\ref{N_d}).
\end{itemize}

\begin{figure}[htbp]
\centerline{\includegraphics[width=\columnwidth]{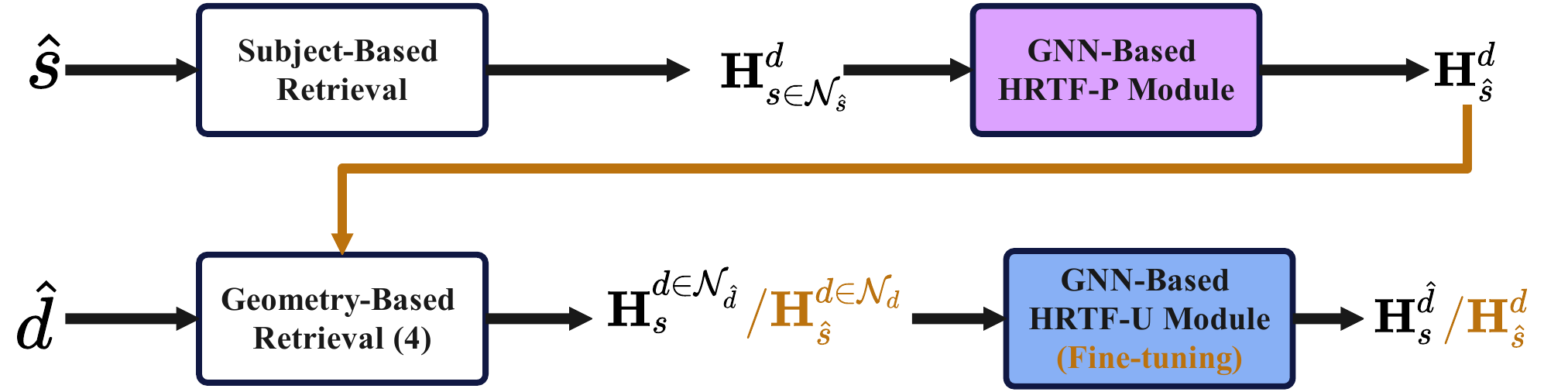}}
\caption{Overview of the proposed HRTF personalization framework (referred to as GraphNF-SCA), where the pre-trained HRTF-U module is fine-tuned to reinforce the spatial correlation among the outputs of the HRTF-P module.
}
\label{fig:overall}
\end{figure}

An overview of the proposed HRTF personalization framework is shown in Figure \ref{fig:overall}, which consists of three core parts: an HRTF-P module, an HRTF-U module, and a fine-tuning stage. 
Specifically, the HRTFs of the subjects retrieved in (\ref{N_s}) are fed into the GNN-based HRTF-P module that is established using the graph-based network architecture (Figure \ref{fig:2}).
At the same time, HRTFs corresponding to the positions retrieved in (\ref{N_d}) are fed into the GNN-based HRTF-U module that is constructed using another GNN-based architecture (Figure \ref{fig:3}).
After pretraining the above two modules, the output $\mathbf{H}_{\hat s}^{d} ( d\in \mathcal{N}_{{d}})$ of the HRTF-P module is used to fine-tune the HRTF-U module, and the latter can generate high-quality personalized HRTFs reinforced by spatial correlations. 
The technical details of each part are described in the following subsections.

\vspace{-0.3cm}

\subsection{GNN-Based HRTF-P Module}
\begin{figure*}[t]
\centering
\includegraphics[width=0.95\textwidth]{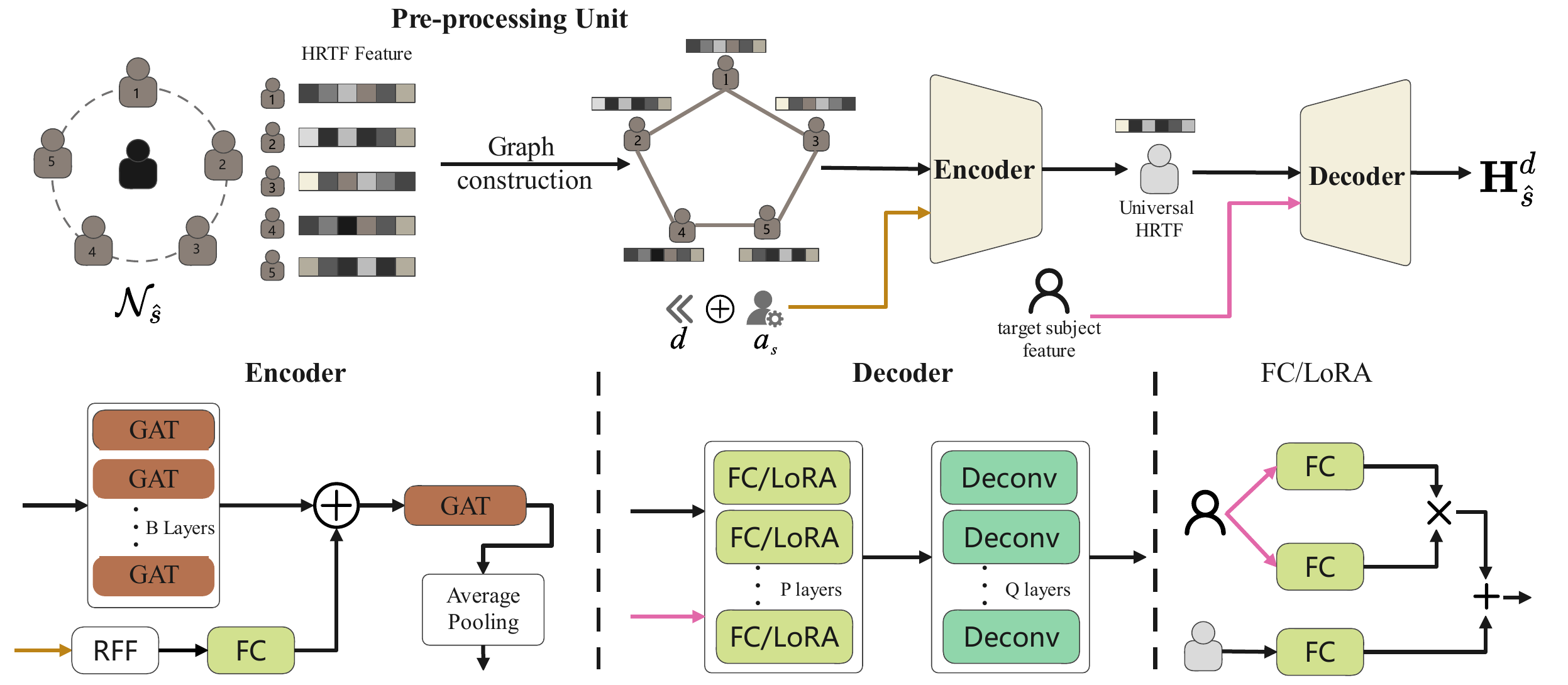}
\caption{Network architecture for GNN-based HRTF-P module.}
\label{fig:2}
\end{figure*}

As shown in Figure \ref{fig:2}, the GNN-based HRTF-P module is composed of a pre-processing unit followed by an encoder-decoder architecture. To be specific, the pre-processing unit prepares inputs for the encoder that extract subject-shared features, while the decoder outputs the personalized HRTFs after embedding subject-specific features.

\subsubsection{Pre-processing Unit}
For a target subject $\hat s$, we first construct a graph $\mathcal{G}_{\hat{s}}$ as
\begin{equation} \label{g_p}
\mathcal{G}_{\hat{s}} = \{ \mathcal{V}_{\hat{s}}, \mathcal{E}_{\hat{s}}, \mathcal{W}_{\hat s} \}
\end{equation}
where $\mathcal{V}_{\hat{s}}$, $\mathcal{E}_{\hat{s}}$, and $\mathcal{W}_{\hat s}$ are the vertex set, the edge set, and the edge-weight set, which are defined by
\begin{subequations} \label{g_vew}
\begin{equation} \label{g_v}
	\mathcal{V}_{\hat{s}} = \{ \mathbf{H}_{s}^{d} \mid s\in \mathcal{N}_{\hat{s}} \}
\end{equation}
\begin{equation}\label{g_e}
    \mathcal{E}_{\hat{s}} =  \{ (s, q) \mid  s, q \in \mathcal{N}_{\hat{s}} \}
\end{equation}
\begin{equation}\label{g_w}
    \mathcal{W}_{\hat s} =   \{ 1 \mid (s, q) \in \mathcal{E}_{\hat{s}} \}\;.
\end{equation}
\end{subequations}
That is, $\mathcal{G}_{\hat{s}} $ is set to a fully-connected graph for mining subject-shared features.

In addition to (\ref{g_p}), we also concatenate the source direction $d$ and the subject-relevant characteristic $a_s$ to build the following clue
\begin{equation}\label{add_Clue}
\mathbf{C}_{s}^{d} = d \oplus a_s ,  \;s \in \mathcal{N}_{\hat{s}}
\end{equation}
with $\oplus$ being the concatenation operator.

\subsubsection{Encoder}\label{Encoder}

The encoder is dedicated to extract the HRTF-relevant feature $\tilde{\mathbf H}_{s}^{d} \in \mathbb{R}^{|\mathcal{N}_{\hat{s}}| \times 2K * N}$ from graph $\mathcal{G}_{\hat{s}}$ and the clue-relevant feature  $\tilde{\mathbf{C}}_{s}^{d} \in \mathbb{R}^{|\mathcal{N}_{\hat{s}}| \times 2K}$ from $\mathbf{C}_{s}^{d}$, then fuse them to output the subject-shared feature $\tilde{\mathbf{F}}^{d} \in \mathbb{R}^{ N \times 2K }$, where $N$ represents the number of attention heads used in the multi-head attention mechanism of the first graph attention (GAT) layer.

The HRTF-relevant feature $\tilde{\mathbf H}_{s}^{d}$ is derived from the vertex embeddings of the graph $\mathcal{G}_{\hat{s}}$ transformed by $B$ successive GAT layers. In each GAT layer, we concatenate $N$ attention heads activated by the Exponential Linear Unit (ELU) function \cite{clevert2015fast}.
Taking the first layer as an example, its output $\tilde{\mathbf H}_{s}^{d}(1)$ can be expressed as
\vspace{-0.2cm}
\begin{align} \label{gat_out}
\tilde{\mathbf H}_{s}^{d}(1)
\!=\! \bigoplus_{n=1}^N \! f_e\!\Big( 
&\alpha_{{s,s}}^{(n)} \mathbf{W}_{s}^{(n)} \tilde{\mathbf H}_{s}^{d}
 \!+\! \sum_{q \in \mathcal{N}_{\hat{s}}} \alpha_{s,q}^{(n)} \mathbf{W}^{(n)} \tilde{\mathbf H}_{q}^{d}
\Big)
\end{align}
where $s \in \mathcal{N}_{\hat{s}}$ is the node index, \( f_e(\cdot) \) represents the ELU activation function, $\mathbf{W}_s^{(n)} \in \mathbb{R}^{M \times 2K}$ and $\mathbf{W}^{(n)} \in \mathbb{R}^{M\times2K}$ are two learnable transformation matrices in the $n$-th attention head with $M$ being the output feature dimension, and  \( \alpha_{s,q}^{(n)} \) represents the attention coefficient between nodes \( s \) and \( q \), i.e.,
\begin{equation}\label{coeff_nodes}
\alpha_{s,q}^{(n)} = f\left[ f_l ( 
\mathbf{a}_{s}^{(n) ^\top} \mathbf{W}_{s}^{(n)} \tilde{\mathbf H}_{s}^{d} + 
 \mathbf{a}^{(n)^ \top} \mathbf{W}^{(n)} \tilde{\mathbf H}_{q}^{d})\right]
\end{equation}
where $\top$ denotes the transpose operator, \( \mathbf{a}_{s}^{(n)} \in \mathbb{R}^{M} \) and \( \mathbf{a}^{(n)} \in \mathbb{R}^{M} \) are two learnable weight vectors, \( f_l(\cdot) \) denotes the LeakyReLU activation function \cite{maas2013rectifier}, and \( f[\cdot] \) represents the Softmax function, which normalizes the computed attention coefficients to facilitate comparison between different nodes.

The clue-relevant feature $\tilde{\mathbf{C}}_{s}^{d}$ is obtained by feeding the random Fourier feature (RFF) \cite{tancik2020fourier} of ${\mathbf{C}}_{s}^{d}$ into a fully connected (FC) layer. Here, the dimension of $\tilde{\mathbf{C}}_{s}^{d}$ is fixed to $2K$ by adjusting the dimension of the FC layer.

The feature fusion is carried out as follows: The  HRTF-relevant feature $\tilde{\mathbf H}_{s}^{d}$ and the clue-relevant feature $\tilde{\mathbf{C}}_{s}^{d}$ are first concatenated to construct a new node feature ${\mathbf{F}}_{s}^{d} \in \mathbb{R}^{|\mathcal{N}_{\hat{s}}| \times (2K * N + 2K)}$. By treating ${\mathbf{F}}_{s}^{d}$ ($s \in \mathcal{N}_{\hat{s}}$) as a vertex set, with edges and weights defined by (\ref{g_e}) and (\ref{g_w}) respectively, we construct a graph processed through a single GAT layer and average pooling, yielding the fused feature $\tilde{\mathbf{F}}^{d}$. After average pooling, the resulting feature $\tilde{\mathbf{F}}^{d}$ eliminates individual differences and is transformed into a universal HRTF that is independent of the subject. Such universal representations can then be reconstructed as personalized HRTFs through the decoder.

\begin{figure}[htbp]
\centerline{\includegraphics[width=\columnwidth]{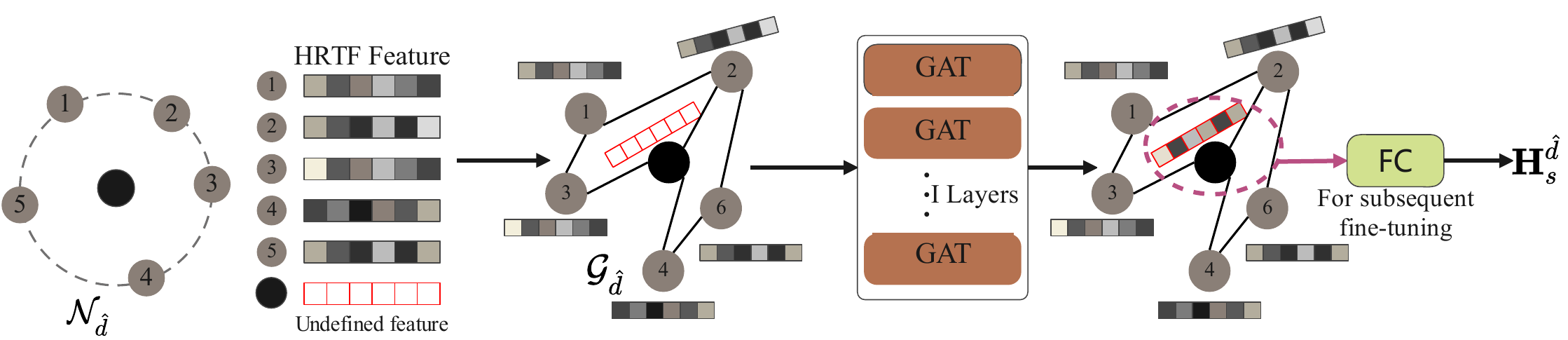}}
\caption{Network architecture for GNN-based HRTF-U module.
}
\label{fig:3}
\end{figure}

\vspace{-0.4cm}
\subsubsection{Decoder} The decoder is used to embed the target subject's feature into the universal representation $\tilde{\mathbf{F}}^{d}$, generating the personalized HRTF ${\mathbf H}_{\hat s}^{d}$. Motivated by \cite{10889481}, we adopt $P$ FC layers with Low-Rank Adaptation (LoRA) \cite{hu2022lora} and $Q$ deconvolution layers to build the decoder. The main difference from \cite{10889481} is that we do not split $\tilde{\mathbf{F}}^{d}$ into HRTF-relevant and clue-relevant parts.
In each layer, the target subject's features are extracted by two FC layers (as depicted in Figure \ref{fig:2}), generating two low-rank matrices \(\mathbf{u} \in \mathbb{R}^{2K \times 1}\) and \(\mathbf{v} \in \mathbb{R}^{2K \times 1}\), respectively. By using these two low-rank matrices, we can adaptively adjust the fused feature \(\tilde{\mathbf{F}}^{d}\) as
\begin{equation}
\bar{\mathbf{F}}^{d} = \bar{\mathbf{F}}^{d} + \mathbf{u} \mathbf{v}^\top
\end{equation}
where \(\bar{\mathbf{F}}^{d}\) is generated by passing \(\tilde{\mathbf{F}}^{d}\) through an FC layer, which has the same dimension with \(\tilde{\mathbf{F}}^{d}\). This adaptive adjustment enables the universal feature \(\tilde{\mathbf{F}}^{d}\) to better match the personalized characteristics of the target subject.

We adopt the log-spectral distortion (LSD) as the loss function to measure the spectral distortion between the predicted and the ground-truth HRTF. The LSD is defined as

\begin{equation}\label{loss_fun}
LSD({\mathbf H}_{\hat s}^{d}, \bar{ \mathbf H}_{\hat s}^{d}) =\sqrt{\frac{1}{2K} \sum\limits_{k = 1}^{2K} \left( 20\log_{10} \frac{\bar{ \mathbf H}_{\hat s}^{d}(k)}{{\mathbf H}_{\hat s}^{d}(k)} \right)^2 }
\end{equation}
where \( {\mathbf H}_{\hat s}^{d}(k)\) represents the \(k\)-th element in ${\mathbf H}_{\hat s}^{d}$, and \( \bar{ \mathbf H}_{\hat s}^{d} \) denotes the true HRTF magnitude for the target subject at direction $d$.

\vspace{-0.2cm}
\subsection{GNN-Based HRTF-U Module}

The GNN-based HRTF-U module is built to learn the spatial correlation among HRTFs. The network architecture is depicted in Figure \ref{fig:3}, and its input is also a graph, termed $\mathcal{G}_{\hat{d}}$, which is determined by
\begin{equation} \label{g_u}
\mathcal{G}_{\hat{d}} = \{ \mathcal{V}_{\hat{d}}, \mathcal{E}_{\hat{d}}, \mathcal{W}_{\hat d} \}
\end{equation}
where $\mathcal{V}_{\hat{d}}$, $\mathcal{E}_{\hat{d}}$, and $\mathcal{W}_{\hat d}$ are the vertex set, the edge set, and the edge-weight set, which are defined by

\begin{subequations} \label{g_dvew}
\begin{equation} \label{g_dv}
	\mathcal{V}_{\hat{d}} = \{ \mathbf{H}_{s}^{d} \mid d\in \mathcal{N}_{\hat{d}} \} \cup  \mathbf{H}_{s}^{\hat d} 
\end{equation}
\begin{equation}\label{g_de}
    \mathcal{E}_{\hat{d}} =  \{ (d, p) \mid \left \| d - p \right \|_2  < a \cdot\delta_d, \; d, p \in \mathcal{N}_{\hat{d}} \}
\end{equation}
\begin{equation}\label{g_dw}
    \mathcal{W}_{\hat d} =   \{\exp(-\frac{\left \| d-p \right \|_2^2}{2 \sigma^2}) \mid (d, p) \in \mathcal{E}_{\hat{d}} \}
\end{equation}
\end{subequations}
where $0 \le a \le 1$ is the factor that controls the graph topology and \(\sigma\) is the bandwidth parameter of the Gaussian kernel function. Unlike (\ref{g_vew}), we use a non-fully connected graph, where the edge and the edge weight are determined by the spatial distance between nodes. This emphasizes the spatial correlation between nodes in the subsequent GNN. In addition, we also put the HRTF $\mathbf{H}_{s}^{\hat d} $ at target position $\hat d$ into the constructed graph in (\ref{g_dv}), which is initialized with all-one vectors. Based on the above, the constructed graph $\mathcal{G}_{\hat{d}}$ is fed into successive $I$ GAT layers and an FC layer. Each GAT layer has the same structure as (\ref{gat_out}) and (\ref{coeff_nodes}). In addition, we use the LSD, which is defined similarly to (\ref{loss_fun}), as the loss function.
Finally, the output is extracted from the target node at the last GAT layer. 

\vspace{-0.1cm}
\subsection{Fine-Tuning}
 During the pre-training stage, we should train the HRTF-P and HRTF-U models in parallel. The former learns the mapping in (\ref{mapp_hrtf_p}), while the latter learns the mapping in (\ref{mapp_hrtf_u}).
After that, for an unseen subject $\hat{s} \notin \mathcal{S}$, we run the HRTF-P module at all positions $d \in \mathcal{D}$ to generate $\mathbf{H}_{\hat s}^{d\in \mathcal{D}}$. Next,  for each position $\hat d  \in \mathcal{D}$, its neighboring points are retrieved via (\ref{g_dv}) to construct the graph (\ref{g_u}), which is then
fed into the pre-trained HRTF-U module to fine-tune the FC layer. In doing so, personalized HRTFs are no longer position-independent after passing through the fine-tuned HRTF-U module. 
Consequently, personalized HRTFs are enhanced through spatial correlations, which is expected to improve accuracy.

\section{Experiments}
\subsection{Datasets and Baseline}
We conducted evaluations on three public HRTF datasets, including the SONICOM \cite{engel2023sonicom} dataset, the CIPIC dataset \cite{algazi2001cipic}, and the HUTUBS dataset \cite{brinkmann2019cross}. Details of these datasets can be found in \textbf{Appendix A}.

To evaluate performance, we compared the proposed method with several existing approaches. For traditional methods, we selected the nearest-neighbor method and the ITD/LSD-based HRTF selection method. For data-driven methods, we included recent advances, such as NF(CbC) \cite{10095801}, NF(LoRA) \cite{10448477}, and RANF \cite{10889481}. Detailed descriptions of these comparison methods can be found in \textbf{Appendix B}.

\subsection{Experimental Setup}
For the HRTF-P module, the first graph attention network in the encoder consists of $B=2$ GAT layers, and the numbers of their attention heads were fixed to 8 and 1 respectively; the second graph attention network in the encoder consisted of a single GAT layer with 6 attention heads; the decoder involved $P=2$ FC layers and $Q=4$ deconvolution layers. The HRTF-U module used a two-layer GAT structure (i.e., $I=2$), which has 8 and 1 attention heads, respectively. For other parameter settings, the Gaussian kernel width \(\sigma\) in  (\ref{g_dw}) was set to 0.5, $a$ and  $\delta_d$  in (\ref{g_de}) were set to 0.75 and 20, respectively.

Regarding the training strategy, we adopted the following multiphase optimization strategy. The pre-training phase of the HRTF-P module used the RAdam optimizer \cite{liu2019variance} with an initial learning rate of 0.001, training for 200 epochs. In addition, a dynamic learning rate decay strategy was employed, which multiplies the learning rate by 0.9 whenever the validation loss does not decrease over 10 consecutive epochs.
The pre-training phase of the HRTF-U module used the Adam optimizer \cite{25diederik2014adam} with an initial learning rate of 0.002, training for 200 epochs. The learning rate was multiplied by 0.95 if the validation loss did not decrease for 3 consecutive epochs. The final fine-tuning phase also used the Adam optimizer with an initial learning rate of 0.002, applying an exponential decay strategy with a decay rate of 0.95 per epoch, and trained for a total of 20 epochs.

\begin{table*}[htbp]
\centering
\caption{LSD, ILD errors for different numbers of measurement directions in LAP challenge 2024.}
\resizebox{\linewidth}{!}{\renewcommand{\arraystretch}{1.1}
\begin{tabular}{cccccccccccc}
\hline
         & \multicolumn{2}{c}{3 measurements} &  & \multicolumn{2}{c}{5 measurements} &  & \multicolumn{2}{c}{19 measurements} &  & \multicolumn{2}{c}{100 measurements} \\ \cline{2-3} \cline{5-6} \cline{8-9} \cline{11-12} 
Methods  & ILD[dB]              & LSD[dB]             &  & ILD[dB]              & LSD[dB]             &  & ILD[dB]              & LSD[dB]              &  & ILD[dB]               & LSD[dB]              \\ \hline
Nearest neighbor  & 7.64              & 8.69             &  & 4.78              & 8.30             &  & 2.99              & 5.42              &  & 1.35               & 3.42             \\

HRTF selection(LSD)  & 1.46              & 5.78             &  & 1.40              & 5.64             &  & 1.54             & 5.47             &  & 1.34               & 5.41              \\

HRTF selection(ITD)  & 1.42              & 6.38           &  & 1.46              & 6.43             &  & 1.55              & 6.71              &  & 1.38               & 6.33              \\ \hline

NF(CbC)& 1.54              & 4.87             &  & 2.04              & 5.22             &  & 1.79              & 5.01              &  & 1.67               & 5.12              \\
NF(LoRA)& 1.28              & 4.73             &  & 1.37              & 4.62             &  & 1.09              & 4.07              &  & 1.06               & 3.83              \\
RANF& 1.21              & 4.41             &  & 1.31              & 4.56             &  & 0.95              & 3.58              &  & 0.76               & 3.04              \\ \hline
GraphNF  & 1.22              & 4.33             &  & 1.30              & 4.53             &  & 0.93                  & 3.51                  &  &  0.76                 & 3.03                 \\
GraphNF-SCA & \textbf{0.96}              & \textbf{3.60}             &  & \textbf{0.91}              & \textbf{3.55}             &  & \textbf{0.78}                 & \textbf{3.12}                  &  &  \textbf{0.70}                  & \textbf{2.72}                 \\ \hline
\end{tabular}
}
\label{tab:LSD/ILD}
\end{table*}

% \vspace{-0.1cm}
\subsection{Evaluation in LAP Challenge 2024}

\begin{figure}[!t]
\centerline{\includegraphics[width=\columnwidth]{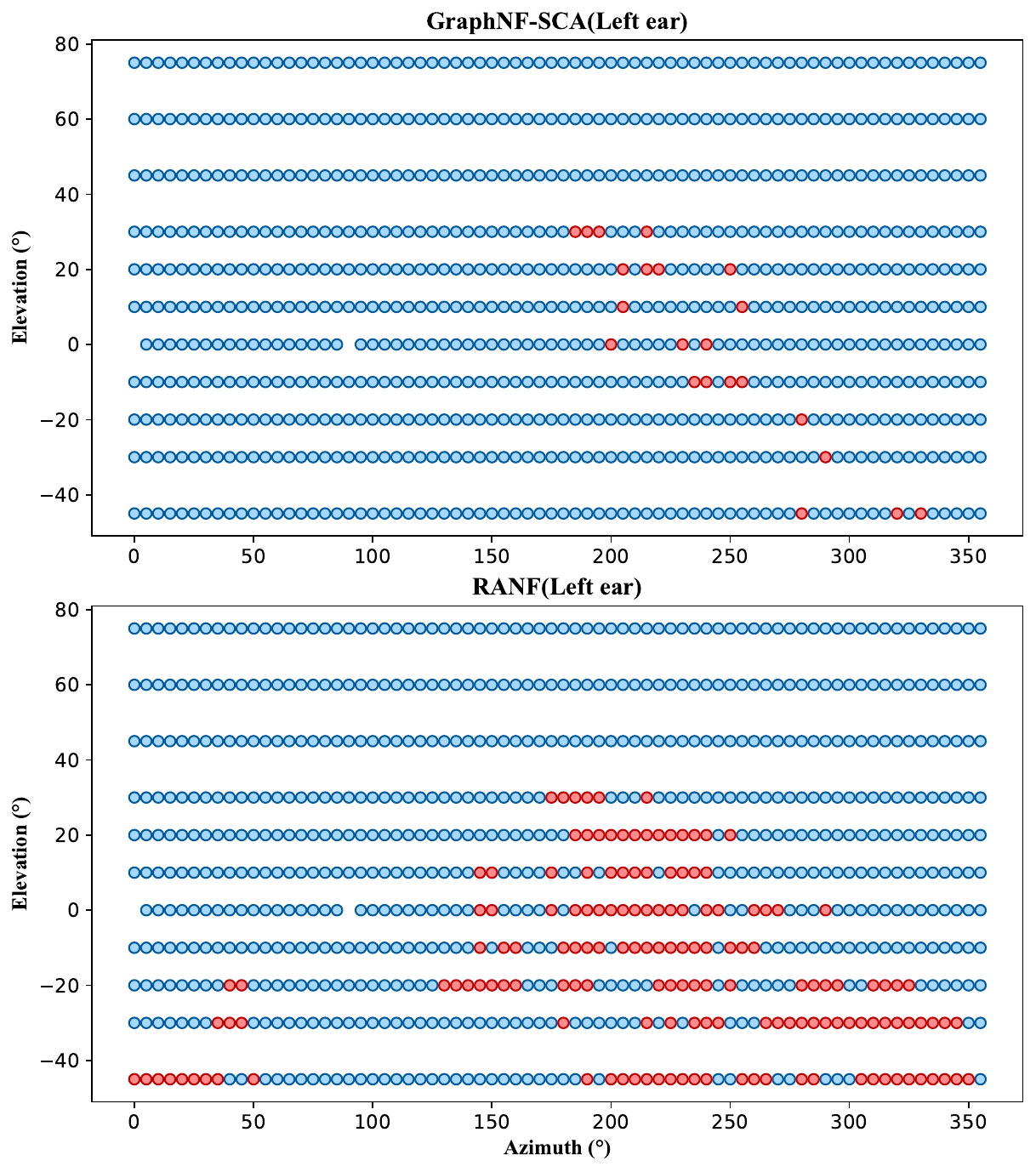}}
\caption{Spatial distribution of LSD errors for unmeasured HRTFs at the left ear. Blue and red dots indicate LSD $\leq \zeta$ and LSD $>\zeta$, respectively.
}
\label{fig:5}
\end{figure}

In this part, we evaluated the HRTF-P performance in the listener acoustic personalization (LAP) challenge 2024 using the SONICOM dataset. 
Due to the fact that the relative relationship between binaural HRTFs fundamentally determines the quality of 3D audio rendering, our evaluation under the LAP Challenge \footnote{https://www.sonicom.eu/lap-challenge} incorporates both the LSD of HRTF magnitude and the interaural level difference (ILD). For the subject-relevant characteristic  $a_s$ in (\ref{N_s}), inspired by the work \cite{10889481}, we use the ILD feature (assuming that the HRTFs of the target subject are measured at a small number of directions in advance) rather than human anatomy. 

In Table \ref{tab:LSD/ILD}, we referred to the HRTF-P module (Figure \ref{fig:2}) as GraphNF, and the augmented version using spatial correlation (Figure \ref{fig:overall}) as GraphNF-SCA.
The ILD and LSD errors were tested under different numbers of HRTF measurements from different directions (3, 5, 19, and 100). 
Overall, the performance of all methods improves as the number of measured HRTF directions increases.
Traditional methods such as Nearest Neighbor and HRTF selection perform poorly under sparse measurement conditions, with LSD errors reaching approximately 6 dB when only 3–5 directions are available.
Data-driven methods outperform traditional approaches, with GraphNF-SCA achieving the best performance.
Specifically, GraphNF slightly outperforms the strongest existing baseline, RANF, indicating that incorporating retrieved subject information via graph-based modeling effectively captures inter-subject relationships and enhances HRTF personalization.
More importantly, GraphNF-SCA consistently achieves the lowest ILD and LSD errors across all configurations.
In an extremely sparse setting with only 3 measurement directions, GraphNF-SCA significantly improves over GraphNF, reducing the LSD from 4.33 dB to 3.60 dB (16.86\%) and the ILD from 1.22 dB to 0.96 dB (21.31\%).
This confirms that the collaborative mechanism between HRTF-P and HRTF-U effectively enhances the performance of HRTF-P and further demonstrates the superiority of GraphNF-SCA in scenarios with severely limited data.

To provide a more intuitive visualization of HRTF estimation results, Figure \ref{fig:5} illustrates the accuracy of HRTF prediction, using the left ear as an example, at all azimuth and elevation angles under the extremely sparse condition of using only 3 measurement directions. The blue and red dots denote HRTFs with errors below and above the threshold $\zeta = 6$ dB, respectively.
It can be observed that LSD errors that exceed the threshold $\zeta$ (at the left ear) are mainly concentrated in the azimuth range of $180^\circ$ to $360^\circ$, corresponding to sound sources located on the right side of the head.
This result aligns with the “acoustic head-shadowing effect”, wherein predicting HRTFs for sources located on the contralateral side results in higher errors due to sound waves being attenuated as they propagate through the head.
Compared to RANF, GraphNF-SCA exhibits a greater number of low-error LSD points (i.e., below the threshold) in unmeasured directions, with high-error regions mainly concentrated on the contralateral side.
To save space, we only present the HRTF estimation performance of RANF and GraphNF-SCA at the left ear. The corresponding results for the right ear, as well as the binaural LSD error distributions of another method, i.e., HRTF Selection (LSD), can be found in \textbf{Appendix C}.

\vspace{-0.1cm}

\begin{table*}[!htbp]
\caption{LSD, ILD errors for different numbers of measurement directions on other datasets.} \resizebox{\linewidth}{!}{\renewcommand{\arraystretch}{1.1}
\begin{tabular}{cccccccccccccc}
\hline
\multirow{2}{*}{Dataset} & \multirow{2}{*}{Methods}              &  & \multicolumn{2}{c}{3 measurements} &  & \multicolumn{2}{c}{5 measurements} &  & \multicolumn{2}{c}{19 measurements} &  & \multicolumn{2}{c}{100 measurements} \\ \cline{4-5} \cline{7-8} \cline{10-11} \cline{13-14} 
                         &                                       &  & ILD{[}dB{]}      & LSD{[}dB{]}     &  & ILD{[}dB{]}      & LSD{[}dB{]}     &  & ILD{[}dB{]}      & LSD{[}dB{]}      &  & ILD{[}dB{]}       & LSD{[}dB{]}      \\ \hline
\multirow{4}{*}{ $\rm{HUTUBS}_{mea}$}  & HRTF selection(LSD)                   &  &1.44                  &5.61                 &  &1.32                  &5.49                 &  &1.33                  & 5.42                 &  & 1.44                  &5.40                  \\
                         & RANF &  & 1.42                 & 4.69                &  &1.24                  &4.43                 &  & 0.93                 &4.20                  &  &0.93                   &3.82                  \\
                         & GraphNF              &  & 1.41              & 4.63             &  & 1.29              & 4.43             &  & 0.91              & 3.94              &  & 0.90               & 3.52              \\
                         & GraphNF-SCA                  &  &\textbf{0.97}                  &\textbf{3.74}                 &  &\textbf{0.93}                  &\textbf{3.72}                 &  &\textbf{0.83}                  &\textbf{3.44}                  &  &\textbf{0.72}                   &\textbf{3.23}                  \\ \hline 
\multirow{4}{*}{ $\rm{HUTUBS}_{sim}$}  & HRTF selection(LSD)                   &  &1.34                  & 4.93                &  &1.23                  &5.21                 &  &1.04                  &5.12                  &  &1.03                   &5.20                  \\
                         & RANF &  &1.24                  &4.13                 &  &1.11                  &4.10                 &  &0.93                  &3.14                  &  &0.83                   &2.77                  \\
                          & GraphNF              &  & 1.22              & 4.12            &  &1.20               &3.94              &  &0.96               &3.15             &  & 0.83               &2.77              \\
                         & GraphNF-SCA                &  &\textbf{0.96}                  &\textbf{3.21}                 &  &\textbf{0.92}                 &\textbf{3.10}                 &  &\textbf{0.81}                  &\textbf{2.74}                  &  &\textbf{0.68}                   &\textbf{2.41}                  \\ \hline
\multirow{4}{*}{CIPIC}   & HRTF selection(LSD)                   &  &2.33                  &6.61                 &  &2.40                  &6.42                 &  &1.81                  &6.34                  &  &1.70                   &6.12                  \\
                         & RANF   & & 1.63              & 4.92                 &  &1.80                 &5.21                 &  &1.59                  &4.32                  &  &1.21                   &3.84                  \\
                         & GraphNF               &  & 1.63              & 4.91             &  & 1.52              & 4.85             &  & 1.44              & 4.21              &  & 1.04               & 3.82              \\
                         & GraphNF-SCA                  &  &\textbf{1.14}                  &\textbf{3.85}                  &  &\textbf{1.13}                   & \textbf{3.83}                 &  &     \textbf{1.07}             & \textbf{3.52}                 &  & \textbf{0.91}                  & \textbf{3.32}                 \\ \hline
\end{tabular}}
\label{table:3}
\end{table*}

\vspace{-0.1cm}

\subsection{Evaluation on Other Databases}

To evaluate the generalization ability of the proposed method on other datasets, we carried out experiments on the following databases: $\rm{HUTUBS}_{mea}$, $\rm{HUTUBS}_{sim}$, and CIPIC. Among them, $\rm{HUTUBS}_{mea}$ and $\rm{HUTUBS}_{sim}$ correspond to the measured and simulated data from the HUTUBS dataset, respectively.
To save space, we only included the best-performing traditional method, namely HRTF selection (LSD).
From Table \ref{table:3}, we observe a trend similar to that in Table \ref{tab:LSD/ILD}: As the amount of measurement data increases, both the ILD and LSD errors decrease accordingly.
Under sparse measurement conditions (3–5 directions), GraphNF achieves an ILD error comparable to the strongest baseline, RANF, while reducing the LSD error by 1.8\%, showing a slight advantage.
Under higher-density measurement conditions (100 directions), GraphNF reduces the ILD error by 4.3\% and the LSD error by 2.1\%, indicating a limited improvement.
Furthermore, by incorporating spatial correlation modeling, GraphNF-SCA demonstrates clear advantages under sparse measurement conditions.
Compared to RANF, GraphNF-SCA achieves a 26.9\% reduction in ILD error and a 23.7\% reduction in LSD error. Even under high-density measurement conditions, it retains a clear advantage, with average reductions of 18.3\% in ILD error and 13.1\% in LSD error. The proposed method consistently outperforms all baselines across all evaluated datasets, with particularly strong performance under extremely data-sparse conditions.

 \vspace{-0.1cm}

\subsection{Ablation Study}
In the ablation study, we first investigated the impact of selecting different subject-relevant characteristics $a_s$  and varying the number $M=|\mathcal{N}_{\hat{s}}|$ of subjects in retrieval (\ref{N_s}), under the extremely sparse condition of using only three measurement directions on the SONICOM dataset.
As shown in Table \ref{table:Ablation}, using the ITD and ILD features for retrieval results in comparable HRTF personalization performance, both slightly outperforming LSD-based retrieval.
In addition, the impact of $M$ is also minor.
Notably, GraphNF-SCA shows a clear and consistent improvement over GraphNF across all settings.

\begin{table}[htbp]
    \centering
    \caption{Impact of different retrieval strategies.}
    \label{tab:Retrieval_Strategy}
    \Large % 调整字体大小
    \resizebox{\linewidth}{!}{ % 让表格适应页面宽度
    \begin{tabular}{cclcclcc}
    \hline
    \multirow{2}{*}{Retrieval} & \multirow{2}{*}{$M$} &  & \multicolumn{2}{c}{GraphNF} &  & \multicolumn{2}{c}{GraphNF-SCA} \\ \cline{4-5} \cline{7-8} 
                                    &                                     &  & ILD{[}dB{]}      & LSD{[}dB{]}      &  & ILD{[}dB{]}        & LSD{[}dB{]}        \\ \hline
    \multirow{3}{*}{LSD}                & 1                                   &  & 1.23              & \textbf{4.30}             &  & 0.94                   &   3.59                     \\
                                        & 5                                   &  & 1.26              & 4.38              &  & 0.94                   &   3.63                   \\
                                        & 10                                  &  & 1.24              & 4.39             &  & 0.93                   &   3.61                    \\ \hline
    \multirow{3}{*}{ITD}          & 1                                   &  & 1.24              & \textbf{4.30}              &  &0.92                    & 3.57                   \\
                                        & 5                                   &  & \textbf{1.22}              & 4.33              &  &0.93                    &  3.61                   \\
                                        & 10                                  &  & 1.23              & 4.39              &  & 0.93                   &   3.64                 \\ \hline
    \multirow{3}{*}{ILD}          & 1                                   &  &1.23              &\textbf{4.30}                  &  &0.93                    &\textbf{3.56}                    \\
                                        & 5                                   &  &\textbf{1.22}              &4.33               &  &0.96                    &3.60                    \\
                                        & 10                                  &  &1.27                  &4.37                  &  &\textbf{0.90}                    & 3.59                   \\ \hline
    \end{tabular}
    }
\label{table:Ablation}
\end{table}

%\vspace{-0.1cm}

\begin{table}[htbp]
    \centering
    \caption{Impact of module integration.}
    \label{tab:Retrieval_Strategy2}
    \begin{tabular}{cccc} % 改成 cccc 表示四列内容居中
    \hline
        & Model & ILD{[}dB{]} & LSD{[}dB{]} \\ \hline
    (a) & GraphNF  (w/o $d$, $a_s$, $g$)
     & 1.69        & 4.65        \\
    (b) &GraphNF  (w/ $d$, $a_s$, w/o $g$)      & 1.34            & 4.41            \\
    (c) &GraphNF        & 1.22        & 4.33            \\
    (d) & GraphNF-SCA      & 0.96        & 3.60            \\ \hline
    \end{tabular}\vspace{-0.10cm}
\end{table}
\vspace{-0.10cm}
The previous experiment settings were kept, and we used the ILD feature to conduct another ablation experiment on the SONICOM dataset.
We compared the following variants of GraphNF: the GraphNF without inputting source direction \(d\) and subject-related feature \(a_s\) (denoted as GraphNF (w/o \(d, a_s\))); the GraphNF incorporating \(d\) and \(a_s\) but without employing the GAT layer for further feature integration (denoted as GraphNF (w/ \(d, a_s\), w/o \(g\))).
As shown in Table \ref{tab:Retrieval_Strategy2}, the HRTF reconstruction error decreases significantly as more modules are integrated into the model.
These results validate the effectiveness of the collaborative mechanism proposed in the GraphNF-SCA network.

\vspace{-0.1cm}
% See Appendix D for further ablation experiments.
For more ablation studies and comparison experiments, please refer to \textbf{Appendix D}.

 \vspace{-0.25cm}
\section{Conclusion}
In this work, we propose GraphNF-SCA, a novel framework for personalized HRTF prediction that effectively incorporates spatial correlation modeling into graph-based learning. Our method consists of three key components: the HRTF-P module, which predicts individual HRTFs via a graph-based encoder-decoder; the HRTF-U module, which captures the spatial structure of HRTFs through a second GNN; and a fine-tuning stage that refines the output of the HRTF-P module by using the fine-tuned HRTF-U module. By explicitly modeling the spatial relationship between HRTFs, GraphNF-SCA significantly outperforms existing methods that perform HRTF personalization independently for each position. Extensive experiments demonstrate that our framework achieves state-of-the-art performance. We believe GraphNF-SCA offers a promising direction for accurate and scalable HRTF personalization in immersive 3D audio applications.

\section{Acknowledgement}
This work was supported by the National Natural Science Foundation of China under Grants 62361045 and 62201297.

% \bibliographystyle{aaai2026}
% \bibliography{aaai2026} 

% \input{./ReproducibilityChecklist.tex}
\newpage
\section{Appendix A}
In this paper, we use the SONICOM, CIPIC, and HUTUBS datasets. The details of these datasets are as follows.

\subsubsection{SONICOM} 
The SONICOM HRTF dataset \cite{engel2023sonicom} comprises multimodal measurement data from 200 subjects, including Head-Related Transfer Functions (HRTFs), Headphone Transfer Functions (HpTFs), depth images, and 3D models of the subjects’ ears, heads, and torsos.
Each subject’s Head-Related Impulse Responses (HRIRs) were measured at 793 spatial positions and sampled at 44.1 kHz, 48 kHz, and 96 kHz.
This study utilizes the 48 kHz HRTF data and excludes subject P0079 due to missing ITD/ILD values.
Regarding dataset partitioning, the first 160 subjects were used for pre-training, the following 19 subjects were used for validation, and the remaining 20 subjects were reserved for evaluation.

\subsubsection{CIPIC} 
The CIPIC HRTF dataset \cite{algazi2001cipic} contains HRTF measurement data from 45 subjects. For each subject, HRIRs were recorded at 1250 spatial positions with a sampling rate of 44.1 kHz. For dataset partitioning, the first 40 subjects were used for pre-training, the following 4 subjects were used for validation, and the remaining 5 subjects were used for evaluation.

\subsubsection{HUTUBS} The HUTUBS dataset \cite{brinkmann2019cross} contains multimodal measurement data from 96 subjects. For each subject, the dataset includes HRIR data at 440 measured positions and 1730 simulated positions. We denote the measured data as $\rm{HUTUBS}_{mea}$ and the simulated data as $\rm{HUTUBS}_{sim}$.
Regarding dataset partitioning, the first 77 subjects were used for pre-training, the following 9 subjects were used for validation, and the remaining 10 subjects were used for evaluation.

\section{Appendix B}
Traditional and data-driven approaches were used as baselines. 
Traditional approaches include the nearest-neighbor algorithm and HRTF selection based on ITD/LSD.
The former selects the HRTF of the spatially closest known sample of the target subject as the estimation result based on spatial distance metrics; the latter determines the optimal reference HRTF from the dataset by calculating the matching error in ITD (Interaural Time Difference) or LSD (Log-Spectral Distortion) between the target subject and reference subject. In other words, the HRTF selection approach operates without requiring HRTF measurements from the target subject.
The data-driven approaches leverage recent advances in Neural Fields (NFs). The NF(CbC) \cite{10095801} method employs a conditioning-by-concatenation strategy, concatenating subject-specific latent vectors with input features to predict the target HRTF. NF(LoRA) \cite{10448477} incorporates Low-Rank Adaptation (LoRA) to fine-tune the fully connected layers using two low-rank matrices, and predicts HRTFs based on azimuth and elevation coordinates. RANF \cite{10889481} also adopts the LoRA mechanism and further integrates ITD-based retrieval to select HRTF from the most similar reference subjects, thus improving HRTF prediction precision.

\section{Appendix C}
\begin{figure*}[!t]
\centering
\includegraphics[width=\textwidth]{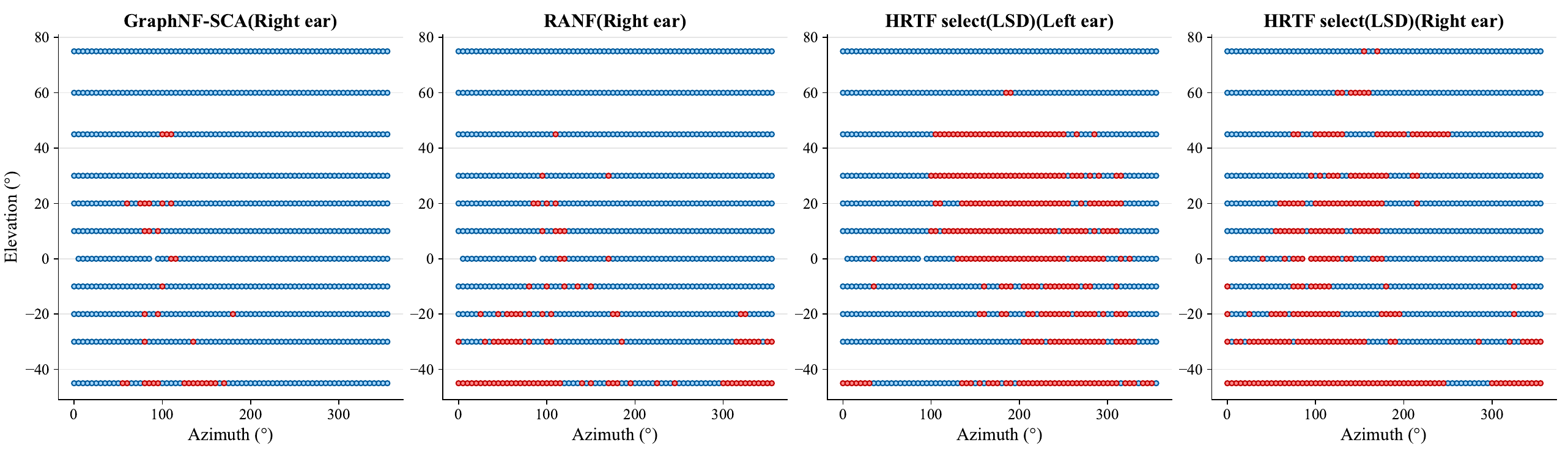}
\caption{LSD error distribution of GraphNF-SCA and RANF at the right ear, 
and the binaural LSD error distributions of HRTF selection (LSD) method. Blue and red dots indicate LSD $\leq \zeta$ and LSD $>\zeta$, respectively (SONICOM HRTF dataset). }
\label{fig:6}
\end{figure*}

In Figure~\ref{fig:6}, we compared the GraphNF-SCA, the RANF, and the LSD-based HRTF selection method under the sparse condition of only three measurement points. Similar to  Figure \ref{fig:5}, we evaluated the prediction performance for unmeasured directions with the threshold $\zeta$ = 6 dB.
The results show that the LSD errors $\geq \zeta$ for the left ear are primarily concentrated in the azimuth range of $180^\circ$–$360^\circ$ (i.e., sound sources on the right side), whereas the right ear exhibits most errors within $0^\circ$–$180^\circ$ (i.e., sound sources on the left side). 
This phenomenon aligns with the acoustic head-shadowing effect, where cross-side prediction (requiring sound waves to traverse the head) results in larger errors than same-side cases.
The HRTF selection method performs the worst, exhibiting widespread LSD errors $\geq \zeta$ in both ears. This confirms that directly using another individual’s HRTF significantly degrades 3D audio rendering effects.
Although the RANF outperformed the HRTF selection method overall, it still exhibits abnormally high errors in certain same-side directions, where low errors would typically be expected.
In contrast, the proposed GraphNF-SCA exhibits only a small number of prediction errors exceeding the threshold in unmeasured directions, all located on the contralateral side.
All same-side predictions yield errors below the $\zeta$ threshold.
These results clearly demonstrate the advantage of our method in mitigating the head-shadowing effect and its superior ability to predict HRTFs accurately in unmeasured directions.

\begin{figure*}[t]
\centering
\includegraphics[width=\textwidth]{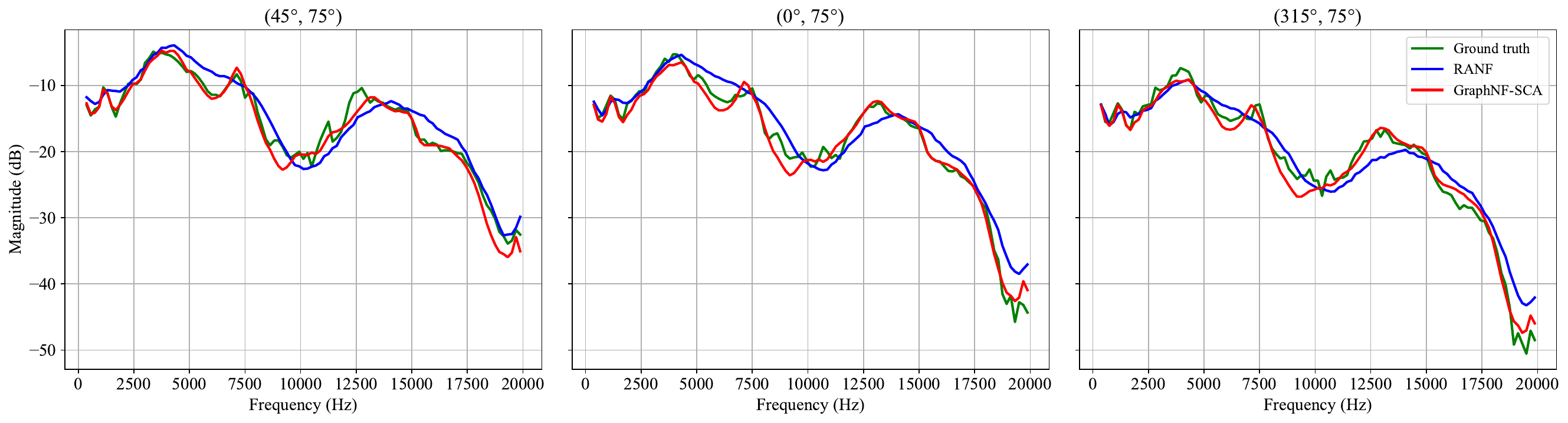}
\caption{HRTF magnitude at the left ear for a subject with 3 different incidence angles (SONICOM HRTF dataset).}
\label{fig:7}
\end{figure*}

\section{Appendix D}
In Table  \ref{table:2}, we evaluated the performance of the HRTF-U module on the HRTF upsampling task. The comparison includes a traditional linear interpolation method and a learning-based approach, FiLM-HRTF \cite{20lee2023global}.
The former estimates HRTFs by interpolating between the two nearest sampling points in the azimuth and elevation planes, while the latter employs a Feature-wise Linear Modulation (FiLM) network to predict  HRTFs using the HRTFs from neighboring sampled points around the target point. In addition, to assess the impact of fine-tuning on the performance of the HRTF-U module, we also compared the model before and after fine-tuning, with the fine-tuned version denoted as $\mathrm{HRTF\text{-}U_{ft}}$.
The results clearly indicate that the traditional linear interpolation method performs the worst across all evaluated datasets.
This is primarily due to its simplistic linear computation, which relies solely on surrounding HRTFs.
In particular, on the HUTUBS and SONICOM datasets—which both employ non-uniform spatial sampling—the interpolation error reaches up to 6 dB.
This degradation can be attributed to the fact that non-uniform sampling severely limits the effectiveness of linear interpolation.
The linear interpolation method performs relatively better on the CIPIC dataset, which adopts a uniform spatial sampling scheme. 
Unlike linear interpolation, the data-driven methods perform robustly across both uniform and non-uniform sampling scenarios.
This can be attributed to the ability of neural networks to learn representations from surrounding points and produce more accurate interpolations at target locations.
Notably, the HRTF-U module outperforms all other methods. 
This superiority stems from the GNN’s ability to adaptively process inputs from a variable number of neighboring points.
In contrast, the fixed-number neighbor sampling strategy used in FiLM-HRTF may miss certain high-information sampling points, leading to instability in the interpolation process.

\begin{table}[!htbp]
    \centering
    \caption{LSD performance in the HRTF upsampling task.}
    \huge % 调整字体大小
    \resizebox{\linewidth}{!}{\renewcommand{\arraystretch}{1.3}
\begin{tabular}{ccccc}
\hline
Method         & $HUTUBS_{sim}$ & $HUTUBS_{mea}$ & SONICOM & CIPIC \\ \hline
Linear Interp. & 6.61  & 6.69 & 6.86 & 3.68  \\
FiLM-HRTF & 1.77  & 3.64 & 2.36 & 3.25  \\
HRTF-U  & \textbf{1.56}  & \textbf{3.18} & \textbf{2.26} & \textbf{3.11} \\ 
$\mathrm{\text{HRTF-U}_{ft}}$  & 1.97  & 3.72 & 2.68 & 3.38\\ \hline
\end{tabular}}
\label{table:2}
\end{table}

It is worth noting that the performance of the fine-tuned model, $\mathrm{HRTF\text{-}U_{ft}}$, is slightly lower than that of the original HRTF-U. This is mainly because the original HRTF-U model had already achieved optimal performance after thorough training, and fine-tuning it introduced certain changes to the model parameters, resulting in a slight decline in performance. Although $\mathrm{HRTF\text{-}U_{ft}}$ sacrifices a certain degree of prediction accuracy in this setting, it significantly improves the overall performance of the HRTF-P module (see Table \ref{table:3}). Moreover, since the HRTF-U module does not require retraining for new users during actual deployment, it ensures the practicality and scalability of the proposed method.

Figure \ref{fig:7} presents the predicted HRTF magnitudes at the left ear of subject P0183 under an extremely sparse measurement condition with only three measurement points, using three representative incidence angles for visualization. In this experiment, we carefully selected three azimuth–elevation combinations: $(0^\circ, 75^\circ)$, $(45^\circ, 75^\circ)$, and $(315^\circ, 75^\circ)$. The elevation angle of $75^\circ$ was chosen to minimize the influence of head geometry on the HRTF—this is quantitatively supported by Figure \ref{fig:5}, where errors at this elevation remain below the threshold $\zeta$. Meanwhile, varying azimuths enable a more comprehensive performance evaluation. The results show that, compared to the state-of-the-art baseline method RANF, the proposed GraphNF-SCA method demonstrates superior performance. Its predicted curves align more closely with the ground-truth HRTFs, capturing spectral peaks and notches more accurately, and maintaining better consistency across the full frequency range. These findings strongly confirm the effectiveness and robustness of GraphNF-SCA in achieving HRTF personalization under extremely sparse measurement conditions.

\end{document}